# Time dilation and length contraction without relativity: The Bohr atom and the semi-classical hydrogen molecule ion

Allan Walstad[*]

**Abstract** Recently it was shown that classical "relativistic" particle dynamics was implicit in physics going back to Maxwell. The demonstration utilized a simple modification of a 1906 thought experiment by which Einstein established the mass equivalence of energy, independently of the relativity postulates. The modified thought experiment leads directly to correct expressions for the momentum and kinetic energy of a particle at high speed. In the present paper, one additional ingredient is added, namely, the Bohr quantization rule of the old quantum theory. It is found that a semi-classical model of a hydrogen atom and a hydrogen molecule ion exhibit, respectively, time dilation and length contraction.

## 1 Introduction

In 1906, Einstein[1] published a thought experiment to establish the mass equivalence of energy, independently of the relativity postulates. A pulse of light is emitted from one end of a box of length $L$ and mass $M$ and absorbed at the other end (see Fig. 1). From Maxwellian electrodynamics, it was known that the light, of energy $E$, would have momentum $p = E/c$. Hence, the box must recoil at speed $V = p/M$ and travel a small distance $\delta = VL/c$ while the light traverses its length. As a result, the center of mass of an isolated system initially at rest would seem to have undergone a spontaneous displacement. This result can be avoided if the energy $E$ carries with it a mass $m = E/c^2$ from one end to the other of the box, to balance the opposite shift of the mass of the box.

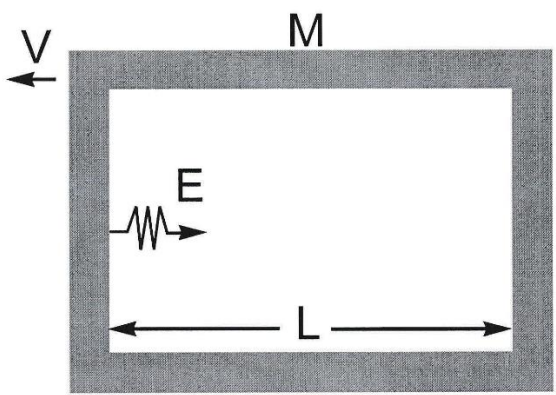

Fig. 1: Einstein's box. A pulse of light having energy E is emitted from one end of a box of length L and mass M. As a result, the box recoils at speed V.

As pointed out in a recent paper[2], by replacing the light pulse with a material particle of mass $m$ (measured at rest), emitted with speed $v$, momentum $p$, and kinetic energy $K$, it is easy to demonstrate via the same reasoning that the momentum of the particle is not $mv$ but

*(m+K/c²)v*. If such a particle is accelerated from rest by a force *F = dp/dt*, setting the work done equal to the kinetic energy yields, after integration, the usual "relativistic" expressions *p = mγv* and *K = m(γ – 1)c²*, where *γ = (1 – β²)¹ᐟ²* is the Lorentz factor and *β* is equal to *v/c*. Thereby we are in possession of relativistic particle dynamics without need of the relativity postulates.

The present paper adds one additional ingredient, namely, the Bohr quantization rule of the old quantum theory. Two simple models are investigated: a Bohr hydrogen atom and a semi-classical model of a hydrogen molecule ion. Traveling at high speed, the former exhibits time dilation of its orbital period, while the latter exhibits length contraction, in accord with the usual relativistic formulas.

## 2 The High-Speed Bohr Atom

Consider a Bohr hydrogen atom traveling along the z axis at speed *v*, with the electron's orbit of radius *r* lying in the *xy* plane (see Fig. 2). According to Maxwellian electromagnetism, the attractive force on the electron is given [see, for example, Becker[3] eq. 64.18] by

$$F = \frac{1}{4\pi\epsilon_0} \frac{e^2}{\gamma r^2},\qquad(1)$$

where $\gamma = \left(1 - \frac{v^2}{c^2}\right)^{-1/2}$ is the Lorentz factor corresponding to the velocity v of the atom. Throughout this paper, the orbital radius r and all other quantities are those observed in the laboratory. No comparison is made between quantities measured by relatively moving observers.

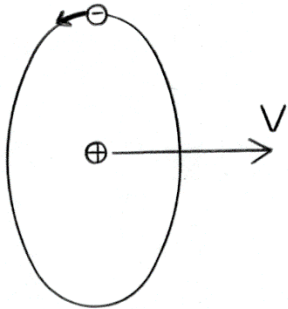

Fig. 2: Bohr atom with velocity perpendicular to the orbital plane.

The electromagnetic attraction accounts for the centripetal force, i.e., the rate of change of momentum of the electron, which is

$$\frac{dp}{dt} = \frac{v_\perp p_\perp}{r},\qquad(2)$$

where $v_\perp$ is the (orbital) velocity component of the electron perpendicular to the z direction and $p_\perp$ is the corresponding momentum component. We know $p_\perp = m\gamma' v_\perp$ where $\gamma'$ is the Lorentz factor associated with the speed of the electron, which is due to the combination of its orbital velocity and the velocity of the atom. Since the angular momentum is $\Lambda = rp_\perp$, the centripetal force becomes

$$F = \frac{\Lambda^2}{m\gamma' r^3}. \tag{3}$$

Using the Bohr quantization rule $\Lambda = n\hbar/2\pi$, where n is a positive integer, we have

$$F = \frac{n^2 \hbar^2}{m\gamma' r^3}. \tag{4}$$

Setting (1) and (4) equal and solving for r, we find

$$r = \frac{4\pi\epsilon_0 n^2 \hbar^2}{e^2 m} \cdot \frac{\gamma}{\gamma'}. \tag{5}$$

It may seem by (5) that the orbital radius as observed in the laboratory depends on the speed of the atom (which would be contrary to the Lorentz transformation in special relativity), but that is not the case. Let us begin with the atom at rest, with the electron of mass m orbiting the nucleus of mass M >> m at speed $v_0$. Its momentum is perpendicular to the z axis and equal in magnitude to $p_0 = m\gamma_0 v_0$. The atom is given an impulse in the z direction sufficient to accelerate it to speed v. Specifically, let the impulse given to the electron be J. As a result the electron acquires the speed v' and momentum p' = m$\gamma$v'. By the composition of vectors we must have $p_0^2 + J^2 = p'^2$ as well as $v/v' = J/p'$. We may eliminate J to obtain

$$\frac{v'^2}{v^2} = 1 - \frac{\gamma_0^2 v_0^2}{\gamma'^2 v'^2}. \tag{6}$$

Using the generic relation $v^2 = (1 - 1/\gamma^2)c^2$ to eliminate v, $v_0$, and v', with a bit of algebra we arrive at the relation

$$\gamma' = \gamma_0 \gamma. \tag{7}$$

Then (5) becomes

$$r = \frac{4\pi\epsilon_0 n^2 \hbar^2}{e^2 m\gamma_0} \tag{8}$$

which is just the expression for the radius of the stationary atom (where γ = 1 and γ′ = γ₀). Hence, the orbital radius (which is perpendicular to the velocity of the atom) does not depend on the atom's speed.

The orbital period is given by

$$T = \frac{2\pi r}{v_\perp}.  \tag{9}$$

Substituting

$$v_\perp = \frac{p_\perp}{m\gamma'} = \frac{\Lambda}{m\gamma' r} = \frac{n\hbar}{m\gamma' r} \tag{10}$$

and utilizing (7) and (8) to eliminate r and γ′, we arrive at

$$T = \frac{32\pi^3 \epsilon_0^2 n^3 \hbar^3 \gamma}{e^4 m \gamma_0}. \tag{11}$$

We see thereby that the orbital period is proportional to the Lorentz factor γ, demonstrating time dilation according to the usual relativistic formula.

### 3 The High-Speed Hydrogen Molecule Ion

We consider now a semi-classical model of a hydrogen molecule ion. The two protons lie along the z axis, separated by a distance $L$ (see Fig. 3). The electron orbits their midpoint in a circle of radius r, with the plane of its orbit lying parallel to the $xy$ plane. The molecule is moving in the z direction at speed $v$, which we again assume is much faster than the electron's $xy$ velocity component $v_\perp$.

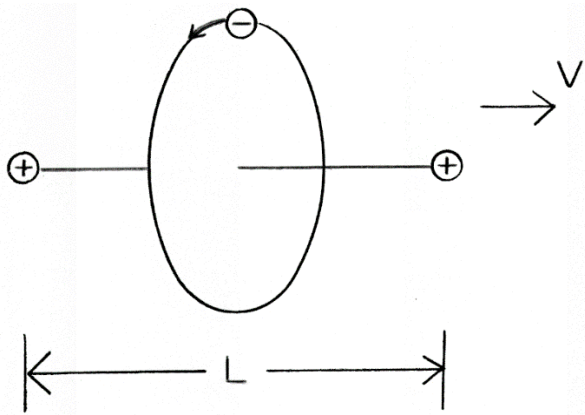

The centripetal force is again given by (4), but the radius of the orbit is now $r = (L/2)\tan\theta$, where $\theta$ is the angle between the z axis and the direction from one proton to the electron. The centripetal force is then

$$F = \frac{8n^2\hbar^2}{m\gamma'L^3\tan^3\theta} . \tag{12}$$

The resultant electromagnetic force on the electron [again via, for example, Becker[3] eq. 64.18] due to the two protons is also toward the center of its orbit and comes out to be

$$F = \frac{e^2}{4\pi\epsilon_0} \frac{8(1-\beta^2)\sin\theta\cos^2\theta}{L^2(1-\beta^2\sin^2\theta)^{3/2}} . \tag{13}$$

We wish to set (9) and (10) equal, but there is one more condition to apply, namely, that the z component of the attractive force of the electron on a proton must equal the repulsive force between the protons, in order to bind the molecule stably. The former force component is [via Becker[3] eq. 64.18]

$$F_z = \frac{e^2}{4\pi\epsilon_0} \frac{(1-\beta^2)\cos\theta}{r_{ep}^2(1-\beta^2\sin^2\theta)^{3/2}} \tag{14}$$

where $r_{ep} = L/2\cos\theta$ is the electron-proton distance.

The proton-proton repulsive force is

$$F_{pp} = \frac{e^2}{4\pi\epsilon_0} \frac{(1-\beta^2)}{L^2} . \tag{15}$$

Equating (11) and (12) leads to the condition

$$(1 - \beta^2\sin^2\theta)^{3/2} = 4\cos^3\theta . \tag{16}$$

Then, equating (9) and (10) and applying (13), it is possible to eliminate θ and solve for

$$L = \frac{4\pi\epsilon_0 n^2\hbar^2}{\left(4^{2/3}-1\right)^2 e^2 m\gamma_0\gamma} . \tag{17}$$

We see thereby that the inter-proton distance is inversely proportional to the Lorentz factor, demonstrating length contraction according to the usual relativistic formula. (For an $H_2^+$ ion at rest, the bond length obtained from this semi-classical analysis is several times smaller than the measured value.)

## 4. Conclusion

There are now several examples of time dilation and length contraction in physical models developed independently of the relativity postulates. The examples presented in this paper utilize the Bohr quantization rule of the old quantum theory.

[*] University of Pittsburgh at Johnstown, Pennsylvania USA